\documentclass[prl,twocolumn,tightenlines,superscriptaddress,preprintnumbers,floatfix,nofootinbib,]{revtex4}

\usepackage{graphicx}
\usepackage{amsmath}
\usepackage{amssymb}
\usepackage{bbold}
\usepackage{color}
\usepackage{bm}
\input epsf

\def\be{\begin{equation}}
\def\ee{\end{equation}}
\def\bea{\begin{eqnarray}}
\def\eea{\end{eqnarray}}

\def\pmax{p_{\rm max}}

\def\p{\mathbf{p}}
\def\k{\mathbf{k}}

\def\ptw{{\tilde{p}}}

\def\ftw{\tilde{f}}

\begin{document}

\title{Approach to equilibrium in weakly coupled nonabelian plasmas}
\author{Aleksi Kurkela}
\affiliation{Physics Department, Theory Unit, CERN, CH-1211 Gen\`eve 23, Switzerland}
\author{Egang Lu}
\preprint{CERN-PH-TH-2014-093}
\affiliation{McGill University, Department of Physics,\\
  3600 rue University, Montr\'eal QC H3A 2T8, Canada}
\begin{abstract}
We follow the time evolution of nonabelian gauge bosons from far-from-equilibrium initial conditions 
to thermal equilibrium by numerically solving
an effective kinetic equation that becomes accurate in the weak coupling limit. We consider initial conditions that are either highly overoccupied or underoccupied. 
We find that overoccupied systems thermalize through a turbulent cascade reaching equilibrium in multiples
of a thermalization time  
$t_{\rm eq}\approx 72./ (1+0.12\log \lambda^{-1}) \times 1/\lambda^2 T$
, whereas underoccupied systems undergo
a ``bottom-up'' thermalization in a time 
$
t_{\rm eq}\approx   \left(34. +21. \ln(Q/T) \right)/ (1+0.037\log \lambda^{-1}) \times \left(Q/T\right)^{1/2}/\lambda^2 T
$, where
$Q$ is the characteristic momentum scale of the initial condition. We apply this result to model 
initial stages of heavy-ion collisions and find rapid thermalization roughly in a time $Qt_{\rm eq} \lesssim 10$ or $t_{\rm eq}\lesssim 1$ fm/c.
\end{abstract}

\maketitle 

\section{Introduction}
Nonabelian far-from-equilibrium plasmas occur in many cosmological pre/reheating scenarios \cite{reheating}
or due to possible cosmological phase transitions \cite{phase_tr}, as well as in the early stages of heavy-ion collisions. 
These far-from-equilibrium systems may be overoccupied, such that the energy is
spread out in longer wave-length modes than in thermal equilibrium but with stronger fields. This is
the case, \emph{e.g.}, for fields generated through parametric resonance, and in
heavy-ion collisions, at least in the limit of asymptotically large centre of mass energies, where the initial condition may be described 
using the color-glass-condensate framework \cite{CGC}. Alternately, far-from equilibrium systems may
be underoccupied, such that the system consists of fewer, but more energetic, quasiparticles than the corresponding thermal system. 
This is the case in, \emph{e.g.}, Planck suppressed decay of inflatons \cite{Cosmo}. Also, even though the initial condition of 
heavy-ion collisions is overoccupied, it has been demonstrated by Baier, Mueller, Schiff and Son \cite{BMSS} (see also \cite{KM2}) that 
the longitudinal expansion renders the prethermal fireball underoccupied before it reaches local thermal equilibrium. 

This has motivated several numerical \cite{Berges:2008mr,Berges:2012ev,Schlichting:2012es,Berges:2013fga,AKML,KM3} 
and analytical \cite{BGLMV, KM1} works to study simple far-from-equilibrium
model systems, in particular that of a single species of gauge bosons in a flat space-time with statistically
isotropic initial conditions at weak coupling, which we investigate in this letter with both over-
and underoccupied initial conditions. In both cases, we follow the time evolution of the system from the initial far-from-equilibrium state to thermal equilibrium and extract the thermalization time, which we define as the exponent governing relaxation of the deviation of the 
first moment of the distribution function from its equilibrium value $\langle |p | \rangle_T$ at late times
\begin{equation} 
\langle |p|(t) \rangle -  \langle |p| \rangle_T \propto e^{-t/t_{\rm eq}}.
\label{t_eq}
\end{equation}

In the overoccupied case, early dynamics fall onto a non-thermal attractor solution: if the initial momentum
scale characterised by $Q^2 \equiv \langle p^2 (t=0)\rangle$
is much smaller than the momentum scale of the target thermal
system, then the scattering time of the initial system $\tau_{\rm init}\sim (Q/T)^7/(\lambda^2 \, T) $  \cite{KM1} is faster than 
that of the thermalized system $\tau_{\rm therm.} \sim 1/(\lambda^2 \, T)$, with $T$ the final equilibrium temperature and
$\lambda \equiv N_c g^2$ the t'Hooft coupling. While it takes the system at least
$\tau_{\rm therm.}$ to reach thermal equilibrium, it only takes $\tau_{\rm init} \ll \tau_{\rm therm.}$ to lose
detailed information of its initial condition. Therefore, at times $\tau_{\rm init} < t < \tau_{\rm therm.}$
the system will reside in a state that is almost independent of the initial condition but time-varying; 
this is the non-thermal fixed point. The fixed point
is described by a scaling form of the occupation function $f(p,t) = t^{-4/7}\tilde{f}(p/p_{\rm max})$,
where the scaling function $\tilde{f}$ is proportional to  $1/p$ at small momenta and decays exponentially above
an evolving scale $p_{\rm max}\propto Q (Qt)^{1/7}$ \cite{KM1,AKML}. There have been several numerical studies demonstrating the approach to and details of the 
fixed point using classical Yang-Mills theory \cite{Berges:2008mr,Berges:2012ev,Schlichting:2012es,KM3} and the effective kinetic theory (EKT) of Arnold, Moore and Yaffe \cite{AMY4,AKML}.
The simulations have so far been limited to the classical limit ($f\gg1$ and $\lambda\ll 1$), and therefore the transition
from the non-thermal fixed point to thermal equilibrium (with $f\sim1$) has not been addressed and current estimates of the thermalization time are purely parametric. 
In this letter we demonstrate using numerical simulations of EKT that the fixed point solution --- and any initial condition that falls to the basin of attraction of the fixed point --- reaches 
thermal equilibrium in time proportional to
\begin{equation}
t^\textrm{over occ.}_{\rm eq}\approx \frac{72.}{ 1+0.12\log \lambda^{-1} } \, \frac{1}{\lambda^2 T}.
\label{eq:over}
\end{equation}

In the underoccupied case, the scattering
time of the initial state is much longer than that of the target thermal system, and therefore the system only loses
information about the initial conditions on final thermalization. But we demonstrate that some features of the intermediate state
do exhibit partial insensitivity to the details of the initial condition. 
The equilibration of the underoccupied system proceeds via ''bottom-up'' thermalization \cite{KM1} (see in 
the case on expanding background \cite{BMSS}): the initial hard 
particles radiate soft radiation, which forms a thermal background through which the hard particles
propagate. Interaction with the thermal background eventually causes the hard particles to undergo a
radiational breakup releasing their energy to the thermal bath. 
Parametrically, the breakup takes place in the same timescale that it takes for a jet of momentum $Q$
to be quenched in a thermal bath of temperature $T$, $\tau_{eq} \sim 1/\lambda^2 T \left( Q/T \right)^{1/2}$.
The dominant interaction is between the hard particles and the thermal bath, not among the hard particles themselves \cite{KM1,BMSS},
and therefore the equilibration depends only on a few averaged features of the initial condition. We find that the creation of the soft thermal
bath is mainly sensitive to the number of particles in the initial condition $n_H=\int_{\mathbf p} f(p)$, whereas the final thermalization time 
\begin{equation}
t^\textrm{under occ.}_{\rm eq}\approx  \frac{34. +21. \ln(Q/T) }{ 1+0.037\log \lambda^{-1}} \, \, \left(\frac{Q}{T}\right)^{1/2}\, \frac{1}{\lambda^2 T}
\label{eq:under}
\end{equation} 
shows sensitivity to the initial momentum scale $Q$.

\section{Effective kinetic theory}
In the weak coupling limit $\lambda \rightarrow 0$, the evolution of modes with perturbative occupancies $ \lambda f(p)\rightarrow0$ and whose momenta are larger the screening scale $p^2 \gg m^2 
  \equiv 2 \lambda \int_{\p}f(p)/|p|$ can be described to leading order in $\lambda f$ by an effective kinetic equation for the color averaged gauge boson distribution function \cite{AMY4}
 \be
 \partial_t f(p,t) = -\mathcal{C}_{2\leftrightarrow 2}[f](p)-\mathcal{C}_{1\leftrightarrow 2}[f](p).
 \ee
 with%
  \footnote{Our matrix element is related to that of \cite{AMY4} by $|\mathcal{M}|^2=\sum_{bcd}|\mathcal{M}^{a b}_{cd}|^2/\nu$, $f = f_a$, and $\gamma = \gamma^{g}_{gg}/\nu$. $\int_{\p}\equiv \int \frac{d^3p}{(2\pi)^3}$}
  \begin{align}
& \mathcal{C}_{2\leftrightarrow 2}[f](p)=  \frac{1}{2} \displaystyle \int_{\k,\p',\k'} \hspace{-0.3cm}\frac{|\mathcal{M}(m)|^2 (2\pi)^4\delta^{(4)}(p+k-p'-k')}{2k \, 2k' \, 2p \,  2p' }\nonumber
\\&\times \{f_p f_k [1+f_{p'}] [1+f_{k'}] -f_{p'} f_{k'}[1+f_{p}] [1+f_{k'}]\}\label{2to2}\nonumber\\
& \mathcal{C}_{1\leftrightarrow 2}[f](p)  =  \frac{(2\pi)^3}{2|p|^2} \displaystyle\int_0^\infty dp' dk' \,  \gamma^p_{p',k'}(m,T_*) \, \nonumber \\
 &\times \{f_p [1+f_{p'}] [1+f_{k'}]-f_{p'} f_{k'} [1+f_{p}]\} \delta (p-p'-k')\nonumber \\
 +& \frac{(2\pi)^3}{|p|^2} \displaystyle\int_0^\infty dp' dk \, \gamma^{p'}_{p,k}(m,T_*) \,  \delta (p+k-p')\\&\times
  \{f_p f_k[1+f_{p'}]-f_{p'} [1+f_{p}] [1+f_{k}]\}. \nonumber
 \end{align}

The effective matrix elements corresponding to elastic scattering ($|\mathcal{M}|^2$) and 
collinear splitting ($\gamma$) have been discussed in detail in refs.~\cite{AMY4, AKML,deepLPM}. 
The elastic collision term includes effects of screening appearing in the leading order by consistently regulating
the Coulombic divergence in $t$ and $u$ channels at the screening scale $m$. The splitting kernel 
includes the effects of Landau-Pomeranchuk-Migdal (LPM) suppression \cite{LPM}  which regulates collinear divergences and depends on $m$
and an effective temperature
\begin{equation}
T_* = \frac{1}{2}\int_{\mathbf{p}}f(p)(1+f(p))\Bigg/\int_{\mathbf{p}}\frac{f(p)}{|p|}
\end{equation}
that is selfconsistently solved (along with $m$) during the simulation.
The effective theory contains no free parameters besides the coupling constant $\lambda$.
 Our numerical implementation is the discrete-$p$ method of \cite{AKML}.

\section{Over-occupied initial conditions}
\begin{figure}
\begin{center}
\includegraphics[width=0.4\textwidth]{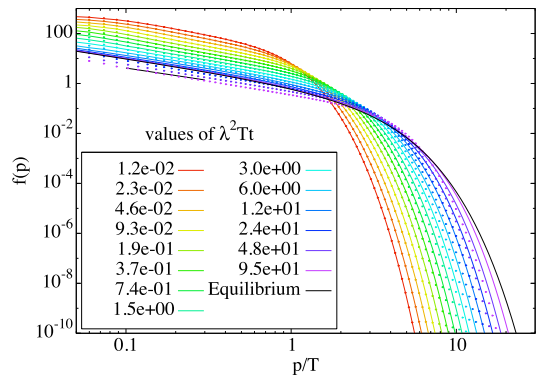}
\caption{ Evolution of $f(p)$ towards the equilibrium
distribution with overoccupied initial conditions. The system starts from the scaling form of Eq.~\ref{eq:scaling} at time $t_0\ll t_{\rm therm.}$ and
relaxes towards the equilibrium distribution denoted by the black line. Even 
at late times the ultraviolet tail of the distribution deviates from the thermal form. The full quantum evolution is compared to the $f\gg 1$ classical approximation well described by Eq.~\ref{eq:scaling} (dotted lines, $y$-axis is $\lambda f$). 
}
\label{f_vs_t}
\end{center}
\end{figure}

\begin{figure}
\begin{center}
\includegraphics[width=0.4\textwidth]{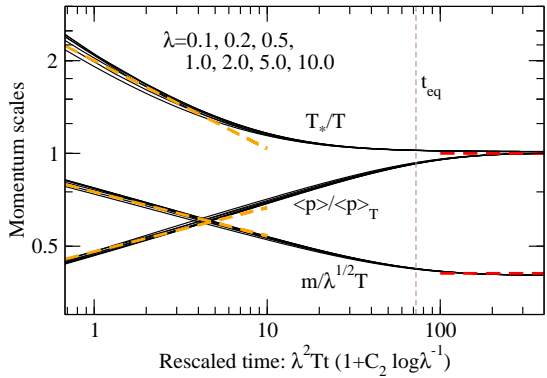}
\caption{Evolution of various moments ($T_*$, $m$ and $\langle |p|\rangle$) as a function of rescaled time with overoccupied initial conditions. 
The different lines correspond to different values of $\lambda=0.1-10$.  The orange dashed lines correspond to the classical limit and the red dashed lines are the thermal values. The vertical line is the thermalization time of Eq.~\ref{eq:over}.
\label{tmp_collapse}
}
\end{center}
\end{figure}

In \cite{AKML}, the form of the fixed point solution of the overoccupied system at times $\tau_{\rm init} \ll t \ll \tau_{\rm therm}$
has been parametrized as $f(p) = (Qt)^{-4/7} \lambda^{-1} \ftw(\ptw)$ with $\ptw =(p/Q) (Qt)^{1/7}$, and
\begin{equation}
 \ftw(\ptw) = \frac{1}{\ptw}\left( 0.22 e^{-13.3 \ptw}+ 2.0 e^{-0.92 \ptw^2 } \right) \label{eq:scaling}.
\end{equation}
This scaling form is expected to be valid as long as the typical momentum scale is small compared to the thermal scale  $\langle | p | \rangle \ll T$, corresponding to $f(\pmax)\gg 1$ and $t \ll \tau_{\rm therm.}$.
If $Q\ll T$, then  $\tau_{\rm init}$ and $\tau_{\rm therm.}$ are parametrically separated, and therefore the system 
will reside in the fixed point for a parametrically long time. Therefore, we can take the fixed point solution as our initial condition
at some time $t_0\ll \tau_{\rm therm.}$ without loss of generality.

Fig.~\ref{f_vs_t} displays the time evolution of the distribution function. The dotted lines are from
a simulation with $\lambda=0$ corresponding to the classical approximation $f\gg 1$. After a quick initial relaxation (not shown in the figure), the distribution function is well described by Eq.~\ref{eq:scaling}. In the classical approximation there is no scale where the cascade would stop and
the cascade continues to move towards the UV \emph{ad infinitum} as described by Eq.~\ref{eq:scaling}. The solid lines correspond to a simulation with the 
same initial $\lambda f$, but with a finite $\lambda = 0.01$. In this case the typical occupancies reach $ f \sim 1$ when $\langle| p| \rangle \sim T$
and the cascade ends to form the equilibrium distribution.

The time evolution of the various relevant moments of the distribution function are shown in Fig.~\ref{tmp_collapse}. 
At times $t \ll \tau_{\rm therm.}$ the moments are well described by power laws predicted in \cite{KM1} (dashed orange lines in the figure) but
when $t \sim \tau_{\rm therm.}$ they smoothly approach their thermal values (red dahsed lines). We repeated the simulation with several $\lambda=0.1-10$. We find that the $\lambda$-dependence is well accounted for by a rescaling of time, $\lambda^2 T t (1+ C_2 \log \lambda^{-1})$ with $C_2=0.12$. The logarithmic dependence in $\lambda$ arises from an IR divergence in momentum diffusion caused by soft elastic collisions in the $m\rightarrow0$ limit.
We show our estimate for the thermalization time of Eq.~\ref{eq:over} with a vertical line, which we have extracted using the definition of Eq.~\ref{t_eq}.
It is noteworthy, however, that the various moments shown in the Figure reach $\sim 50\%$ of their thermal values within a much faster timescale $\sim 0.01t_{eq}$.

\section{Bottom-up thermalization}

\begin{figure}
\begin{center}
\includegraphics[width=0.45\textwidth]{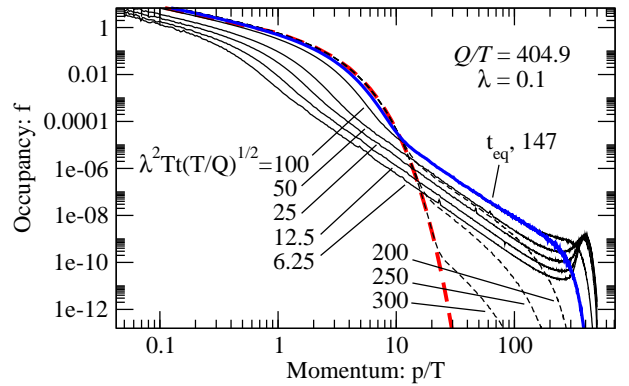}
\caption{ The time evolution of Run 2. At early times soft radiation emitted by the hard particles
creates a thermal bath (solid lines). When the hard particles have had time to undergo a democratic
splitting to daughters of comparable energy, the hard particles undergo a radiational breakup (dashed lines) 
and become part of the thermalized bath. The solid blue line corresponds  to the thermalization time of Eq.~\ref{eq:under} while the red dashed line is the equilibrium distribution. 
}
\label{f_vs_p_2}
\end{center}
\end{figure}

For the under-occupied system the choice of initial conditions will affect the system 
until it is thermalized and therefore we do not experience similar universal behaviour
as in the overoccupied case.  Here, we will proceed by first choosing a specific set of initial 
conditions. For maximally different initial conditions, we use either a step function or a gaussian profile for the distribution function
\begin{equation}
f_{\rm step}(p) \propto \Theta (Q_s- p ), \quad
f_{g}(p) \propto \exp\left[ -\frac{(Q_s- p)^2}{(Q_s/10)^2} \right]
\label{init}
\end{equation}
and fix the constant of proportionality so that the final temperature $T$ is in each case
is the same. 
Our simulation parameters are given in Table \ref{table2}.

The full time evolution of a representative initial condition from $t=0$ to thermalization is shown in figure~\ref{f_vs_p_2}.
Two distinct stages are seen in the evolution: at early times (solid lines) the initial hard particles at the
scale $Q$ emit soft radiation leading to a build-up of a soft thermal bath. At late times (black dashed lines) the hard particles
undergo radiational break-up and become part of the thermal bath (red dashed line).

\begin{figure}
\begin{center}
\includegraphics[width=0.45\textwidth]{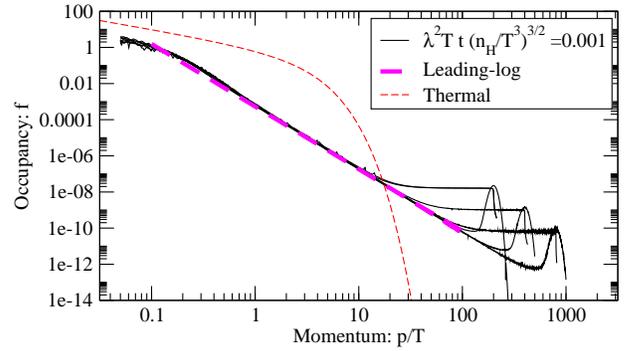}
\caption{The distribution function of various underoccupied initial conditions (gaussian and step function) each shown
after a small fixed amount of rescaled time $t$. 
The dashed magenta and red lines correspond to Eq.~\ref{fs} and equilibrium distribution, correspondingly.  }
\label{creation of soft particles}
\end{center}
\end{figure}

At early times there are three clearly separated structures visible in the spectrum. (i)  The hard particles residing
at the initial scale $Q$ have not had time to scatter and remain very close the initial condition. (ii) The hard particles
emit LPM suppressed radiation with a characteristic $f\propto p^{-7/2}$ spectrum. (iii) In the far IR where $f\gtrsim 1$, reinteractions are fast enough to bring soft modes close to a thermal form. 
The LPM suppressed spectrum can be extracted from the analytically known high energy (leading-log) limit of the
splitting function $\gamma$ \cite{deepLPM}
\begin{align}
\partial_t f(p) &  
\approx  \frac{(2\pi)^3}{p^2} \int_0^{\infty} dp_h f(p_h)\lim_{p\rightarrow 0}\gamma^{p_h}_{p,p_h-p}(T_*,m)  \label{fs} \\
f(p)& \approx  \frac{1}{\sqrt{8\pi}}\frac{n_H}{p^{7/2}} \left(  \lambda^{3/2}\sqrt{T_* m^2}t\right)\log^{1/2}\left[\left(\frac{\lambda}{\pi} \frac{T_* p}{m^2}\right)^{1/2}\right] \nonumber
\end{align} 
At early times, the hard particles dominate the dynamics and hence also $T_*$ and $m^2$. Then 
we can estimate the combination $T_* m^2$ by $\lambda n_H$ and see that the LPM suppressed spectrum, and therefore also the soft thermal bath, are sensitive (up to a log) to the initial conditions at early times only through
the initial number of hard particles $n_H$ --- each hard particle acts only as a source of radiation.  
Figure \ref{creation of soft particles} displays
a mix of Gaussian and step-cutoff simulations (with $\lambda=0.1$), each shown freeze-frame after a small amount
of time. The time is chosen such that the combination $\lambda^2 n_H^{3/2}t$ appearing in Eq.~\ref{fs} is kept constant
between the runs. The good collapse of the distribution functions demonstrates the insensitivity of the 
IR to the details of the momentum distribution in the UV. 

The underoccupied system finally thermalizes through a radiative breakup. Once the
hard particles have had time to undergo a single democratic splitting, \emph{i.e.}, splitting to
daughters with comparable momenta, the resulting
daughters will split again in a time scale that is faster than the initial splitting \cite{BMSS,Turb}.
The daughters undergo successive resplittings until they have deposited their energy into the thermal bath (dashed lines in Figure \ref{creation of soft particles}).
Parametrically, the rate for the first hard splitting and therefore for the cascade is 
\begin{equation}
\tau_{\rm split} \sim (Q/T)^{1/2}\times 1/\lambda^2 T 
\label{tsplit}
\end{equation} up to logarithmic corrections \cite{KM1}.
As in the case of the overoccupied system, we observe that the dependence of the various moments of distribution function on $\lambda$ can be accounted for by a multiplicative rescaling of the time variable, including a logarithm $\lambda^2 T t (1+C_3\log \lambda^{-1})$ with $C_3\approx0.037$. We furthermore find that most of the dependence on the initial condition can be accounted for by rescaling the time by $(T/Q)^{1/2}$ as expected from the parametric estimate for the thermalization time.

\begin{figure}
\begin{center}
\includegraphics[width=0.4\textwidth]{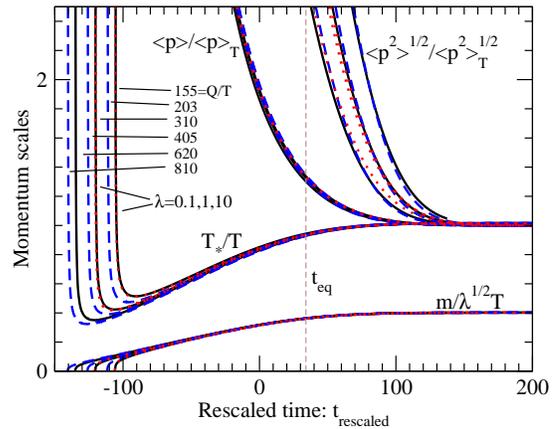}
\caption{Evolution of various moments of the distribution function from the simulations of Table \ref{table2} as a function of rescaled time, Eq.~\ref{tres}.
The dashed blue lines correspond to Gaussian initial conditions with $\lambda=0.1$ whereas solid black ($\lambda=0.1$) and dotted red lines ($\lambda=1,10$, on top of black lines) have step-cutoff initial conditions. $Q$ decreases from left to right. The dashed vertical line is the the thermalization time as given by Eq.~\ref{eq:over}.}
\label{m_vs_t}
\end{center}
\end{figure}

 After the rescaling, $m$, $T_*$, and $\langle |p| \rangle$ show only weak dependence on the initial conditions which is well described by a logarithmic delay of the thermalization as a function of $Q/T$: we find a satisfactory collapse of the data at late times when plotted as a function of 
 \begin{equation}
 t_{\rm rescaled}=\lambda^2 T t (1+C_3\log \lambda^{-1})(T/Q)^{1/2}-C_4\log(Q/T)
 \label{tres}
 \end{equation}
 with $C_4\approx 21$ as seen in Fig.~\ref{m_vs_t}.
We extract the thermalization time from exponential decay of $\langle | p |\rangle-\langle |p| \rangle_T$ as a function of the rescaled 
time variable giving $t_{\rm eq}|_{\rm rescaled}\approx 34.$ (vertical line in the Figure). Converting this back to unscaled time leads
to the thermalization time of  Eq.~\ref{eq:under}.
 We observe, however, that high moments (such as $\langle p^2\rangle$) show sensitivity to the initial conditions arising from increased sensitivity to harder parts of the distribution function, and the dependence to the initial conditions is not removed by a simple rescaling of time.
Unlike in the overoccupied case, the system (in particular the high moments) differs significantly from the equilibrium system before the thermalization time.

\begin{table}
\begin{tabular}{c|cccc||c|cccc}
run &$ Q/T$ & $n_H/n_T$ & $\lambda$ & init &
run &$ Q/T$ & $n_H/n_T$ & $\lambda$ & init \\
\hline
1 & 202.5 & 0.0134 & 0.1 & g  & 
4 & 	155.1 & 0.01799 & 0.1 &step \\
2 & 404.9 & 0.00668 & 0.1 & g  &
5 & 	310.0 & 0.00900& 0.1 &step \\
3 & 809.8	& 	0.00334 & 0.1 & g &
6 & 	620.0 & 0.00450& 0.1 &step \\
\hline 
7 & 	155.137 & 0.01799 & 1.0 &step &
9 & 310.0 & 0.00900& 1.0 &step\\
8 & 	155.137 & 0.01799 & 10.0 & step & 
10 & 310.0 & 0.00900& 10.0 &step \\
\end{tabular}

\caption{Table of simulation parameters for the underoccupied system. The initial particle number density is given in units of the thermal density $n_T$. The last column refers to the initial condition as given by Eq.~\ref{init}.}
\label{table2}
\end{table}
\section{Summary and conclusions}
In this paper, we have for the first time simulated the time evolution of weakly coupled 
nonabelian plasmas from far-from-equilibrium initial conditions to thermal equilibrium. 
We performed the simulation in a framework of an effective kinetic theory that has no free 
parameters and gives an accurate description of the gauge theory in the combined
limit of $\lambda f \ll 1$ and $\lambda \ll 1$ \cite{AMY4}. 

How soon the system can be considered thermal is observable 
dependent. Here we have chosen to define the thermalization time
through the relaxation of mean momentum, but defining the thermalization time
through the relaxation of $m^2$ or $T_*$ results in values compatible with our definition.

Our result can be applied to elevate the parametric weak coupling estimate
of thermalization time in heavy-ion collisions of Baier, Mueller, Schiff and Son to a numerical one \cite{BMSS}.
Their estimate can be quickly derived by 
assuming that as a result of longitudinal expansion the energy density of the system falls 
as a function of (proper) time, $\epsilon \sim \frac{Q^3}{\lambda t}$; so the target temperature falls as $T^4 \sim Q^{3}/ \lambda t$. Inserting this in to
the parametric estimate of the thermalization time of an underoccupied system of Eq.~\ref{tsplit},
and solving for the thermalization time by equating $\tau_{eq}\sim t$ gives 
the estimate of \cite{BMSS}: $Q t \sim \alpha_s^{-13/5}$. 
Replacing the parametric estimate for the thermalization time by Eq.~\ref{eq:under}, 
assuming that the energy density is given by $\epsilon\approx 1.5 d_A Q^4/\pi \lambda (Qt)$ 
 (with $d_A=8$) \cite{Lappi}, choosing $\lambda=11$ corresponding to $\alpha_s \approx 0.3$ and solving the thermalization time selfconsistently gives $Qt_{eq} \approx 1.5$. 
Smaller values of $\alpha_s$ lead to slower thermalization: for $\alpha_s = 0.2$ we get $Q t_{\rm eq}\approx 4.0$. Varying the estimate for the energy density by a factor of 2 gives $Qt_{\rm eq} \lesssim 2.5$ ($Qt_{\rm eq} \lesssim 8$ for $\alpha_s=0.2$) whereas multiplying $\epsilon$ by $(Qt)^{-1/3}$ to estimate an upper bound for redshift effects gives
$Qt_{\rm eq} \lesssim 4$ (for $\alpha=0.3$). For $Q\sim 2$GeV these values correspond to an early thermalization time of $t_{eq}\approx 0.2-1\,$fm/c.  Our conclusion is that rapid thermalization is not in contradiction with weak coupling picture.

The estimate for the thermalization time can be further improved by taking into account the angular dependence of the distribution
function. However, the dominant interaction causing the radiative breakup of the hard particles is against 
the thermal (and hence nearly isotropic \cite{KM2}) background, and therefore we believe that introducing
angular dependence to the EKT will not change the result qualitatively. We also expect 
minor corrections from including fermions in the EKT. We leave these improvements for a forthcoming publication.

\section{Acknowledgements}
The authors thank Juergen Berges, Simon Caron-Huot, Mathias Garny, Fran{\c c}ois Gelis, Liam Keegan, Tuomas Lappi, Aleksi Vuorinen, and Urs Wiedemann for useful discussions.
The authors are indebted to Guy Moore for support, advice, and collaboration in the development of some of the tools used in this study.


\begin{thebibliography}{}
\bibitem{reheating}
  L.~Kofman, A.~D.~Linde and A.~A.~Starobinsky,
  Phys.\ Rev.\ Lett.\  {\bf 73} (1994) 3195
  [hep-th/9405187].
  
  J.~Garcia-Bellido, D.~G.~Figueroa and J.~Rubio,
  Phys.\ Rev.\ D {\bf 79} (2009) 063531
  [arXiv:0812.4624 [hep-ph]].
  
  K.~Enqvist, S.~Nurmi and S.~Rusak,
  arXiv:1404.3631 [astro-ph.CO].

\bibitem{phase_tr}
  J.~M.~Cline,
  hep-ph/0609145.

\bibitem{CGC}
  E.~Iancu, A.~Leonidov and L.~McLerran,
  hep-ph/0202270.
 
  E.~Iancu and R.~Venugopalan,
  In *Hwa, R.C. (ed.) et al.: Quark gluon plasma* 249-3363
  [hep-ph/0303204].
  
  F.~Gelis, T.~Lappi and R.~Venugopalan,
  Int.\ J.\ Mod.\ Phys.\ E {\bf 16} (2007) 2595
  [arXiv:0708.0047 [hep-ph]].

  F.~Gelis, E.~Iancu, J.~Jalilian-Marian and R.~Venugopalan,
  Ann.\ Rev.\ Nucl.\ Part.\ Sci.\  {\bf 60} (2010) 463
  [arXiv:1002.0333 [hep-ph]].


\bibitem{Cosmo}

  K.~Harigaya, M.~Kawasaki, K.~Mukaida and M.~Yamada,
  arXiv:1402.2846 [hep-ph].

  K.~Mukaida and K.~Nakayama,
  JCAP {\bf 1303} (2013) 002
  [arXiv:1212.4985 [hep-ph]].


  K.~Harigaya and K.~Mukaida,
  arXiv:1312.3097 [hep-ph].




\bibitem{BMSS}
  R.~Baier, A.~H.~Mueller, D.~Schiff and D.~T.~Son,
  Phys.\ Lett.\ B {\bf 502} (2001) 51
  [hep-ph/0009237].

 
\bibitem{KM2}
  A.~Kurkela and G.~D.~Moore,
  JHEP {\bf 1111} (2011) 120
  [arXiv:1108.4684 [hep-ph]].

 
\bibitem{Berges:2008mr}
  J.~Berges, S.~Scheffler and D.~Sexty,
  Phys.\ Lett.\ B {\bf 681} (2009) 362
  [arXiv:0811.4293 [hep-ph]].
 
\bibitem{Berges:2012ev}
  J.~Berges, S.~Schlichting and D.~Sexty,
  Phys.\ Rev.\ D {\bf 86} (2012) 074006
  [arXiv:1203.4646 [hep-ph]].
   
\bibitem{Schlichting:2012es}
  S.~Schlichting,
  Phys.\ Rev.\ D {\bf 86} (2012) 065008
  [arXiv:1207.1450 [hep-ph]].
  
\bibitem{KM3}
  A.~Kurkela and G.~D.~Moore,
  Phys.\ Rev.\ D {\bf 86} (2012) 056008
  [arXiv:1207.1663 [hep-ph]].

\bibitem{Berges:2013fga}
  J.~Berges, K.~Boguslavski, S.~Schlichting and R.~Venugopalan,
  arXiv:1311.3005 [hep-ph].

\bibitem{AKML}
  M.~C.~Abraao York, A.~Kurkela, E.~Lu and G.~D.~Moore,
  arXiv:1401.3751 [hep-ph].
  
\bibitem{KM1}
  A.~Kurkela and G.~D.~Moore,
  JHEP {\bf 1112} (2011) 044
  [arXiv:1107.5050 [hep-ph]].

\bibitem{BGLMV}
  J.~-P.~Blaizot, F.~Gelis, J.~-F.~Liao, L.~McLerran and R.~Venugopalan,
  Nucl.\ Phys.\ A {\bf 873} (2012) 68
  [arXiv:1107.5296 [hep-ph]].
  
   
  \bibitem{AMY4}
  P.~B.~Arnold, G.~D.~Moore and L.~G.~Yaffe,
  JHEP {\bf 0301} (2003) 030
  [hep-ph/0209353].



\bibitem{deepLPM}
  P.~B.~Arnold and C.~Dogan,
  Phys.\ Rev.\ D {\bf 78} (2008) 065008
  [arXiv:0804.3359 [hep-ph]].

\bibitem{LPM}
  L.~D.~Landau and I.~Pomeranchuk,
  Dokl.\ Akad.\ Nauk Ser.\ Fiz.\  {\bf 92} (1953) 535.
  %
 A.~B.~Migdal,
  Phys.\ Rev.\  {\bf 103} (1956) 1811.
  %
    R.~Baier, Y.~L.~Dokshitzer, A.~H.~Mueller, S.~Peigne and D.~Schiff,
  Nucl.\ Phys.\ B {\bf 484} (1997) 265
  [hep-ph/9608322].

\bibitem{Turb}
  J.~-P.~Blaizot, E.~Iancu and Y.~Mehtar-Tani,
  Phys.\ Rev.\ Lett.\  {\bf 111} (2013) 052001
  [arXiv:1301.6102 [hep-ph]].

\bibitem{Lappi}
  T.~Lappi,
  Phys.\ Lett.\ B {\bf 703} (2011) 325
  [arXiv:1105.5511 [hep-ph]].

   \end{thebibliography}
\end{document}